\documentclass[a4paper,12pt,english]{article}
\usepackage[cp1251]{inputenc}
\usepackage[T2A]{fontenc}
\usepackage[english]{babel}
\usepackage[dvips]{graphicx}
\usepackage{amsmath}
\usepackage{amssymb}
\usepackage{amsxtra}
\usepackage{amsthm,amscd}
\usepackage{hyperref}

\usepackage{geometry}
\newcommand {\bR}{\mathbb{R}}
\def\mU{\mathfrak{A}}
\def\ma{\alpha}
\def\mb{\beta}
\def\mo{\omega}
\def\mX{\mathfrak{X}}
\def\mR{\mathfrak{R}}
\def\ch{\mathop{\rm ch}\nolimits}

\begin{document}

\title{Characteristic class of a bundle\\ and the existence of a global Routh function\footnote{Translated from \textit{Funktsional'nyi Analiz i Ego Prilozheniya}, Vol. 11, No. 1, pp. 89-90, January-March, 1977. Original article submitted May 14, 1976.
}}

\author{M.P.\,Kharlamov\footnote{Moscow State University}}

\date{}

\maketitle

\begin{center}
{\bf \textit{Functional Analysis and Its Applications},\\
January–March, 1977, Volume 11, Issue 1, pp 80-81 }

\vspace{8mm}

\href{http://link.springer.com/article/10.1007\%2FBF01135548}{http://link.springer.com/article/10.1007\%2FBF01135548}

\vspace{8mm}

\href{http://dx.doi.org/10.1007/BF01135548}{http://dx.doi.org/10.1007/BF01135548}

\end{center}

\begin{abstract}
The possibility of the global Lagrangian reduction of a mechanical system with symmetry is shown to be connected with the characteristic class of a principal fiber bundle of the configuration space over the factor manifold. It is proved that the reduced system is globally Lagrangian if and only if the product of the momentum constant with this characteristic class is zero. In the case of a rigid body rotating about a fixed point in an axially symmetric force field the bundle over a 2-sphere is non-trivial, therefore the reduced system admits a global Routh function if and only if the momentum constant is zero.
\end{abstract}

The invariant analog of natural systems with a cyclic coordinate is given by mechanical systems on manifolds admitting a symmetry group [1] isomorphic to the group $S^1=\{z\in \mathbb{C}:|z|=1\}$. Under certain assumptions concerning the action of $S^1$ on the configuration space $M$ of the original system, a local procedure of the classical Routh method [2] determines a dynamic system on the tangent space to the manifold $M/S^1$ and this system is Lagrangian in some neighborhood of each point. Below we show that the conditions for the existence in such a system of a global Lagrange function of natural form are determined exclusively in terms of the algebraic properties of the foliation of the manifold $M$ into the orbits of $S^1$.

The main object is a mechanical system with symmetry $(M, \mu, V, S^1)$, where $M$ is
a smooth $(n + 1)$-dimensional manifold, $\mu$ is a Riemann metric on $M$, $V$ is a smooth function on $M$, and $S^1$ acts on $M$ without fixed points, preserving $\mu$ and $V$. The system's dynamics is determined by the Lagrange vector field $X$ on $T(M)$ generated by the Lagrange function $L(w)= (\mu(w,w)/2)+V \circ \tau_M(w)$. By $T$ we denote the tangent functor; $\tau_M: T(M) \to M$ is a projection. The manifold $M$ is the total space of a smooth principal $S^1$-bundle $\omega$ with a base $B$ and a projection $p: M \to B$. Let $\mU$ be the maximal set of coordinate neighborhoods $U_\ma$ of the manifold $B$ such that for each $U_\ma\in \mU$ there exists a local trivialization $\Phi_\ma: p^{-1}(U_\ma)\to U_\ma{\times}S^1$ of the bundle $\mo$. If $u = (u^1,\ldots,u^n)$ is a coordinate system in $U_\ma$, then putting $\zeta=(2\pi)^{-1}\mathop{\rm Arg}\nolimits (p_2\circ\Phi_\ma)$, where ${p_2: U_\ma{\times}S^1 \to S^1}$ is the projection onto the second factor, we obtain a system of local coordinates $(u,\zeta)$ in the open set $p^{-1}(U_\ma)$. The collection of functions $(u,\zeta,\dot u,\dot \zeta)$, where $\dot u^i = du^i$, $\dot \zeta^i = d\zeta^i$, introduces local coordinates on the set $\tau^{-1}_M(p^{-1}(U_\ma))$. The function $L$ written in such coordinates is independent of $\zeta$ by virtue of the $S^1$-invariance, and, consequently, $\zeta$ is a cyclic (ignorable) coordinate of the original system. The cyclic integral $\partial L/\partial \dot \zeta$ is a restriction to $\tau^{-1}_M(p^{-1}(U_\ma))$ of the integral $J$ corresponding to the symmetry group by Noether's theorem [3]. Let us fix a value $k \in \bR$. The set $J_k = J^{-1}(k)$ is an $S^1$-invariant submanifold in $T(M)$ (see [1]). The map $\rho: J_k \to T(B)$ coinciding with the restriction to $J_k$ of the map $T(p): T(M) \to T(B)$, gives the factorization of $J_k$. According to Routh's statement, the map $\rho$ takes each  integral curves of the field $X$ lying in the domain $\tau^{-1}_M(p^{-1}(U_\ma))\cap J_k$ to the integral curves of the Lagrange vector field $\mX_\ma$ on $T(U_\ma)$ with the Lagrange function $R_\ma$ defined by the relation
\begin{equation}\label{eq1}
  R_\ma \circ \rho(u,\zeta,\dot u,\dot \zeta)=L(u,-,\dot u,\dot \zeta)-k \dot \zeta.
\end{equation}
Note that on each fiber of $T(U_\ma)$ the function $R_\ma$ is a second-degree polynomial, where the second-degree terms form a nondegenerate quadratic form. A function $F$ of such a form is said to be quadratic and its quadratically homogeneous part is denoted $Q(F)$.

Let $U_\mb \in \mU$, $v= (v^1,\ldots,v^n)$ be a coordinate system in $U_\mb$, $(v, \eta)$ be the corresponding coordinate system in $p^{-1}(U_\mb)$, and $U_\ma \cap U_\mb \ne \varnothing$. The coordinate transformation on the set $p^{-1}(U_\ma \cap U_\mb)$ has the form
\begin{equation}\label{eq2}
  u^i=u^i(v), \qquad \zeta=\eta+(2\pi)^{-1}\mathop{\rm Arg}\nolimits h_{\ma \mb}(v),
\end{equation}
where $h_{\ma \mb}: U_\ma\cap U_\mb \to S^1$ is the transition function of the charts $(U_\ma, \Phi_\ma)$ and $(U_\mb,\Phi_\mb)$ of the bundle $\mo$. The same procedure as above leads to the vector field $\mX_\mb$ on $T(U_\mb)$ with the Lagrange function $R_\mb: T(U_\mb) \to \bR$ satisfying the relation analogous to \eqref{eq1}. Using \eqref{eq2} we obtain that, in the domain $\tau_B^{-1}(U_\ma\cap U_\mb)$,
\begin{equation}\label{eq3}
  R_\mb-R_\ma = (2\pi)^{-1} k\, d \mathop{\rm Arg}\nolimits h_{\ma \mb}.
\end{equation}

\vspace{2mm}

{\bf Lemma}. {\it Let $L_1$ and $L_2$ be quadratic functions on the tangent space $T(N)$ of a connected manifold $N$ and let $Q(L_1) = Q(L_2)$. The Lagrange vector fields on $T(N)$ generated by the functions $L_1$ and $L_2$ coincide if and only if there exist a closed $1$-form $\lambda$ on $N$ and a constant $c$ such that $L_1(w)-L_2(w)= \lambda(w)+c$, $w \in T(N)$.}

\vspace{2mm}

By applying the assertion of this lemma to the functions $R_\ma$ and $R_\mb$, from \eqref{eq3} we get that in the common domain of definition $\mX_\ma=\mX_\mb$ and, consequently, we can define a vector field $\mX(k)$ on $T(B)$ by setting $\mX(k) = \mX_\gamma$ on the set $\tau^{-1}_B(U_\gamma)$ for any $U_\gamma\in \mU$. In addition, the function $K$ is well defined coinciding on each $\tau^{-1}_B(U_\gamma)$ ($U_\gamma\in \mU$) with the function $Q(R_\gamma)$.

Similar to the Hamiltonian case [3], the field $\mX(k)$ is called the reduced system and the function $K$ is called the kinetic energy of the reduced motion. Let us find out the cases when the reduced system globally admits a quadratic Lagrange function with the kinetic energy $K$.

\vspace{2mm}

{\bf Definition}. {\it Let $R$ be a quadratic function on $T(B)$ such that $Q(R) = K$ and $\mX(k)$ coincides with the Lagrange vector field generated by the function $R$. Then $R$ is called the global Routh function of the reduced system $\mX(k)$.}

\vspace{2mm}

Let $\{U_\ma\}$ be a covering of $B$ by the sets from $\mU$. The collection $\{(2\pi)^{-1} d \mathop{\rm Arg}\nolimits h_{\ma \mb}\}$ is a one-dimensional cocycle of the covering $\{U_\ma\}$ with coefficients in the sheaf $\mR$ of germs of closed 1-forms on $B$. The cohomology class of this cocycle determines an element $\chi(\mo)$ of the group $H^1(B,\mR)$ not depending on the original covering.

\vspace{2mm}

{\bf Theorem}. {\it A global Routh function of the reduced system $\mX(k)$ exists if and only if
${k\,\chi(\mo) = 0}$.}

\vspace{2mm}

Let us note that the set of classes of isomorphic (in differentiable sense) principal $S^1$-bundles with the base $B$ is in one-to-one correspondence with the group $H^2(B, \mathbb{Z})$ of the integral cohomologies. Suppose that the bundle $\mo$ is defined by the cocycle $\{U_\ma,h_{\ma\mb}\}$, where $\{U_\ma\}$ is a covering of $B$ by the sets from $\mU$, and the intersection of any two of these sets is simply connected. Let $\{h_{\ma\mb}\}$ be the corresponding system of transition functions. Then the element of $H^2(B, \mathbb{Z})$ corresponding to the bundle $\mo$ is generated by the cocycle $\{n_{\ma\mb\gamma}\}$, where $2\pi n _{\ma\mb\gamma}= \mathop{\rm Arg}\nolimits h_{\mb \gamma}-\mathop{\rm Arg}\nolimits h_{\ma \gamma}+\mathop{\rm Arg}\nolimits h_{\ma \mb}$.
This element is called \textit{the characteristic class} of $\mo$ and is denoted $\ch(\mo)$. Considering any integral cocycle as a real one we get the canonical homomorphism $\nu: H^2(B,\mathbb{Z}) \to H^2(B, \bR)$. The image of $\ch(\mo)$ under this homomorphism and under the de Rham isomorphism $H^2(B,\bR)\to H^1(B,\mR)$ coincides with $\chi(\mo)$ (see [4]).

Thus, when $\nu$ is a monomorphism, i.e., the group $H^2(B, \mathbb{Z})$ does not contain nonzero periodic elements, the reduced system $\mX(k)$ possesses a global Routh function with a nonzero momentum constant $k$ if and only if the bundle $\mo = (M, p, B)$ is differentiably trivial.

As an example we consider the problem of the motion of a rigid body about a fixed
point in a force field with a potential having a symmetry axis passing through the fixed
point. The system's configuration space is $M = SO(3)$, the symmetry group $S^1$
acts as the group of rotations around the symmetry axis, and the factor space $SO(3)/S^1$ is the
two-dimensional sphere $S^2$. The group $H^2(S^2,\mathbb{Z}) = \mathbb{Z}$, the bundle $\mo = (SO(3), p, S^2)$ is isomorphic to the nontrivial bundle of unit tangent vectors to the two-dimensional sphere, and, consequently, $\chi(\mo)\ne 0$. Therefore for a nonzero momentum constant the reduced system in this problem does not have a global Routh function.

The author thanks V. M. Alekseev for attention to the work.

\vspace{5mm}

\end{document}